\documentclass[a4paper, 12pt]{article}

\usepackage{arxiv}

\usepackage[utf8]{inputenc} 
\usepackage[T1]{fontenc}    
\usepackage{hyperref}       
\usepackage{url}            
\usepackage{booktabs}       
\usepackage{amsfonts,amsthm,amssymb,mathtools}       
\usepackage{nicefrac}       
\usepackage{microtype}      
\usepackage{lipsum}		
\usepackage{graphicx,tikz}

\usepackage{doi}
\usepackage{cleveref}

\renewcommand{\vec}[1]{\boldsymbol{#1}}

\DeclareMathOperator*{\argmin}{arg\,min}

\newcommand{\Wtau}{W_{\vec\tau}}
\newcommand{\We}{W_{\mathbf e}}
\newcommand{\sigmatau}{\vec \sigma_{\vec \tau}}
\newcommand{\sigmatauhat}{\hat{\vec \sigma}_{\vec \tau}}

\newcommand{\Eh}{\mathbf{E}}
\newcommand{\tr}{\mathrm{tr}}

\newcommand{\bb}{\mathcal B}

\usepackage[most]{tcolorbox} 

\newtcolorbox[auto counter]{optionalnote}[2][]{
    parbox=false,
    colbacktitle= white,
    colback=green!5!white,
    colframe=white!45!black,
    coltitle=black,
    enhanced,
    attach boxed title to top left={yshift=-1mm},
    title={\thetcbcounter.~#2}
,#1}

\newtcolorbox{highlight-result}[1][]{
 parbox=false,
 boxrule=0pt,top=0pt,bottom=0pt,
colback=blue!8!white,
enhanced,#1}

\tcbset{colback=blue!8!white, colframe=white!50!white,top=0mm,bottom=0mm,right=0mm,left=0mm}

\title{The elastic stored energy of initially strained, or stressed, materials: restrictions and third-order expansions}

\author{ \href{https://orcid.org/0000-0000-0000-0000}{\includegraphics[scale=0.06]{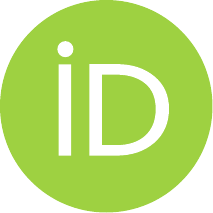}\hspace{1mm}Artur L. Gower}\thanks{webpage:\url{https://arturgower.github.io/}} \\
	Department of Mechanical Engineering\\
	University of Sheffield\\
	United Kingdom \\
	\texttt{arturgower@gmail.com} \\
	\And
     \href{https://orcid.org/0000-0000-0000-0000}{\includegraphics[scale=0.06]{orcid.pdf}\hspace{1mm}Tom Shearer} \\
	Department of Mathematics\\
	 University of Manchester\\
	 United Kingdom\\
	 \texttt{tom.shearer@manchester.ac.uk} \\
  \And
     \href{https://orcid.org/0000-0002-1011-5587}{\includegraphics[scale=0.06]{orcid.pdf}\hspace{1mm}Pasquale Ciarletta} \\
	MOX Laboratory, Department of Mathematics\\
	 Politecnico di Milano\\
  Italy\\
	 \texttt{pasquale.ciarletta@polimi.it}
  	\And
     \href{https://orcid.org/0000-0000-0000-0000}{\includegraphics[scale=0.06]{orcid.pdf}\hspace{1mm}Michel Destrade} \\
	School of Mathematical and Statistical Sciences\\
	 University of Galway\\
	 Republic of Ireland\\
	 \texttt{michel.destrade@universityofgalway.ie}
}

\date{}

\renewcommand{\shorttitle}{\textit{arXiv} Template}

\hypersetup{
pdftitle={Elasticity with initial stress},
pdfauthor={Artur L.~Gower},
pdfkeywords={Nonlinear elasticity, constitutive theory, initial stress, third-order elasticity},
}

\begin{document}
\maketitle

\begin{abstract}
A large variety of materials, widely encountered both in engineering applications and in the biological realm, are characterised by a non-vanishing internal stress distribution, even in the absence of external deformations or applied forces. These initial stresses are due to initial strains or microstructural changes, such as thermal expansion, manufacturing processes, volumetric growth, remodelling, etc. 
A common constitutive choice for modelling such materials extends the classical approach in the field theory of continuum mechanics to include, explicitly, the initial stress or strain in the elastic stored energy function. Here, we discuss why these energy functions need to satisfy some restrictions to avoid unphysical behaviours, such as non-conservation of energy, {and we derive the required restrictions from the classical assumptions of elasticity.} To illustrate their need, we perform a rigorous asymptotic expansion for proving that these restrictions on stored energy functions that depend on the strain and initial stress are required for consistency with strain energy functions of classical third-order weakly nonlinear elasticity. 
\end{abstract}



\section{Introduction}


The classical approach to model nonlinear elastic materials in the field theory of continuum mechanics is to define a strain energy density function, $W=W(\mathbf F)$, where $\mathbf F$ is an elastic deformation gradient from the unstressed reference configuration to the current configuration. The functional form of $W(\mathbf F)$ depends upon the symmetry group corresponding to the underlying material microstructure \cite{gurtin1982introduction, ogden1997non}. In many materials, however, some level of stress is \textit{always} present. This is especially the case in living materials, where mechano-transduction is know to regulate both biochemical activities and gene expression  from cell to tissue level \cite{ambrosi2008stress,ciarletta2016residual}, and in manufactured solids, following thermal changes, fatigue, plasticity, etc. For such materials, allowing $W$ to depend \textit{only} on $\mathbf{F}$ is no longer appropriate.

One constitutive choice to model how an initial stress $\vec \tau$ (the Cauchy stress tensor in the reference configuration) affects the elastic response of a material is to consider a strain energy function of the form $W(\mathbf F \mathbf F_1)$, where both $\mathbf F$ and $\mathbf F_1$ are elastic deformation gradients: $\mathbf F_1$ represents a deformation from a stress-free (possibly virtual) configuration to the configuration with stress $\vec \tau$, and $\mathbf F$ represents a further deformation from the stressed reference configuration to the current configuration. This method, adapted by Rodriguez \textit{et al.} \cite{rodriguez1994stress} from plasticity theory \cite{bilby1956continuous,kroner1959allgemeine}, is called the \textit{multiplicative decomposition method} (see \cite{sadik2017origins} for a historical account). This method works particularly well when the deformation tensor $\mathbf F_1$ that led to the stress $\vec\tau$ is known {\it a priori}.

An alternative way to account for the presence of stress in the reference configuration is to consider a stored energy function $\Wtau(\mathbf F, \vec \tau)$ which explicitly depends on the Cauchy stress tensor $\vec \tau$ in the reference configuration (to our knowledge, Johnson and Hoger \cite{johnson1995use} were the first to introduce this double dependence for the stress response). 
In this paper, we use the term ``stored energy function" instead of strain energy function to emphasise that such functions do not depend \textit{solely} on the strain. In \cite{gower2017new}, we established that functions of the form $\Wtau(\mathbf F, \vec \tau)$ need to satisfy an additional restriction compared to the classical elastic strain energy functions. We called this restriction \textit{Initial Stress Reference Independence} (ISRI), and specific cases of this restriction for stored energy functions that depend \textit{only} on $\mathbf F$ and $\vec \tau$ were derived \cite{gower2015initial,gower2017new}, with further improvement for computational applications in the incompressible limit \cite{magri2024modelling}. 

Currently, there is not a consensus on whether such restrictions are necessary \cite{ogden2023change}. We prove in this work that these restrictions are necessary, starting from simple motivating examples. We first prove that stored energies of the form $W(\mathbf F, \mathbf F_1)$ have to satisfy a restriction that is analogous to the ISRI restriction for  $\Wtau(\mathbf F, \vec \tau)$, relying only on the classical assumptions of nonlinear elasticity, as laid out by Marsden and Hughes~\cite{marsden1994mathematical}, for example. In essence, the restriction on $W(\mathbf F, \mathbf F_1)$ states that the stored energy can only depend on $\mathbf F$ and $\mathbf F_1$ through the term $\mathbf F \mathbf F_1$, which makes these stored energy functions equivalent to the multiplicative decomposition method. From this restriction, we later prove corresponding restrictions on $\Wtau(\mathbf F, \vec \tau)$ by inverting the relation between $\vec \tau$ and $\mathbf{F}_1$.

The article is structured as follows. In \Cref{sec:elastic-spring}, we discuss a simple introductory example to illustrate the necessity of restricting the potential energy of a linear elastic spring under an initial force. In \Cref{sec:strained energies}, we provide a rigorous derivation for a restriction on strain energy functions that depend on both an initial deformation gradient and a subsequent deformation gradient, starting from the classical  assumptions of the field theory of nonlinear elasticity.  This provides a direct way to state that these restrictions are natural and necessary. In \Cref{sec:initially stress energies}, we deduce similar restrictions on stored energy functions that depend on an initial \textit{stress} and a subsequent deformation gradient. Finally, in \Cref{sec:third-order energies}, we prove that restrictions on stored energy functions that depend on the strain and initial stress up to third-order are required to make them consistent with classical third-order elasticity. 


\section{An introductory example}
\label{sec:elastic-spring}

\begin{figure}[h]
    \centering
    \includegraphics[width=0.5\linewidth]{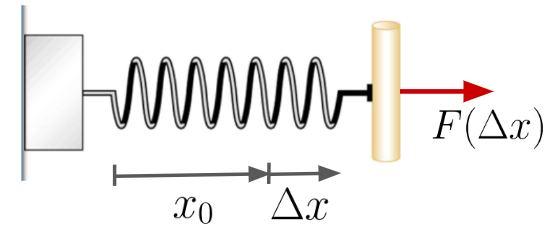}
    \caption{Illustration of an elastic spring with natural length $x_0$, adapted from \cite{conceptualphysics}. As the extension $\Delta x$ increases, the force $F(\Delta x)$ exerted onto the spring also increases. The question answered in this section is: \textit{how should the potential energy depend on $\Delta x$ and the initial force $\tau$ when the spring is initially extended by $\Delta x_\tau$?}}
    \label{fig:spring}
\end{figure}

As an illustrative example, let us consider a one-dimensional elastic spring of natural length $x_0$, aligned with the $X$-axis, with left side fixed at $X=0$ and right side initially at $X=x_0$, which defines the \emph{natural length} of the spring at rest, in the absence of any load. 
The potential energy stored in this spring, when extended to length $x$, is
\begin{equation} \label{eqn:spring energy}
    U(\Delta x) = \tfrac{1}{2}k (\Delta x)^2,
\end{equation}
where $k>0$ is the spring constant and $\Delta x = x - x_0$ is the extension, see \Cref{fig:spring}  for an illustration. 
As the spring is elastic, we can derive from the principle of energy conservation  that the force exerted onto the right side of the spring to maintain the extension $\Delta x$ is
\begin{equation} \label{eqn:spring stress}
    F(\Delta x) = \frac{\mathrm dU(\Delta x)}{\mathrm d \Delta x} = k \Delta x.
\end{equation}

Instead of measuring the energy from the position of rest, we can measure it from a position in which the spring has already been extended by an amount $\Delta x_\tau$. In this position, the spring is subject to the initial force
\begin{equation} \label{eqn:spring stress-linear}
    \tau=F(\Delta x_\tau) = k \Delta x_\tau,
\end{equation}
and we can write the total energy stored in the spring after a subsequent extension $\Delta x$ as a function, $U_\tau$, of both $\Delta x $ and the initial force, $\tau$:
\begin{equation} \label{eqn:spring initial stress to classical}
    U_\tau(\Delta x,\tau) := U(\Delta x + \Delta x_\tau) = \tfrac{1}{2}k(\Delta x + \tau/ k)^2.
\end{equation}
Again, due to the principle of energy conservation, the force exerted onto the spring is 
\begin{equation}\label{eqn: force residual}
F_\tau(\Delta x,\tau)=\frac{\mathrm dU_\tau(\Delta x,\tau)}{\mathrm d\Delta x}=k(\Delta x+\tau/k)=k(\Delta x+\Delta x_\tau)=F(\Delta x+\Delta x_\tau).
\end{equation}
Using two superposed displacements  ($\Delta x_\tau$ followed by $\Delta x$) is a common way to couple force and displacement (or, equivalently, stress and strain, when considering two- and three-dimensional problems). 

An alternative approach is to \emph{choose} a functional form for the stored energy function directly. For example, we could consider the following expression:
\begin{equation} \label{eqn:spring guess}
    \hat{U}_\tau(\Delta x,\tau) = \tfrac{1}{2}k(\Delta x)^2 + \tau\Delta x,
\end{equation}
which leads to the same force as before:
\begin{equation} \label{eqn:spring guess stress}
    \hat{F}_\tau(\Delta x, \tau) = 
    \frac{\mathrm d \hat{U}_\tau(\Delta x, \tau)}{\mathrm d \Delta x} = k \Delta x + \tau=k(\Delta x+\Delta x_\tau)=F(\Delta x+\Delta x_\tau).
\end{equation}
Both equations \eqref{eqn: force residual} and \eqref{eqn:spring guess stress} predict the correct \textit{forces} in the reference position, as $F_\tau(0, \tau)=\hat{F}_\tau(0, \tau) = \tau$, and current position, as $F_\tau(\Delta x,\tau)=F(\Delta x+\Delta x_\tau),~\hat{F}_\tau(\Delta x,\tau)=\hat{F}(\Delta x+\Delta x_\tau)$. However, the amount of energy that is predicted to be stored by equation \eqref{eqn:spring guess} is not consistent with classical spring elasticity, since $\hat{U}_\tau(\Delta x,\tau)\ne \hat{U}(\Delta x + \Delta x_\tau)$. This simple example shows that  choosing a form for the stored energy function requires a great care. 
Even choosing a form that gives the correct force is not sufficient, and our choice is restricted.

One may argue that, although equation \eqref{eqn:spring guess} is not consistent with \textit{classical} spring elasticity, perhaps it is a \textit{self}-consistent constitutive model for an initially stressed spring; however, this is not the case. To emphasise this point, we now show that equation \eqref{eqn:spring guess} predicts different amounts of energy are stored by the same deformation depending on which configuration is chosen as the reference configuration, whereas equation \ref{eqn:spring initial stress to classical} predicts the same amounts. 

If we use equation \eqref{eqn:spring guess} from the stress-free reference configuration (i.e. $\tau=0$) to predict the energy stored by the two extensions $\Delta x_\tau$ and $\Delta x+\Delta x_\tau$, we obtain
\begin{equation}
\hat{U}_\tau(\Delta x_\tau,0) = \frac{k}{2} \Delta x_\tau^2,\quad\hat{U}_\tau(\Delta x+\Delta x_\tau,0) = \frac{k}{2} (\Delta x+\Delta x_\tau)^2,
\end{equation}
If, instead, we choose the configuration associated with the initial extension $\Delta x_\tau$ as the reference configuration, so that $\tau=\hat{F}_\tau(\Delta x_\tau,0)=k\Delta x_\tau$, then equation \eqref{eqn:spring guess} predicts
\begin{equation}
\hat{U}_\tau(\Delta x,F_\tau(\Delta x_\tau,0))=\frac{k}{2} \Delta x^2 + k\Delta x_\tau\Delta x\ne \hat{U}_\tau(\Delta_x+\Delta x_\tau,0).
\end{equation}
The left and right sides of this equation are both predictions of the energy stored in the spring as a result of the same extension $\Delta x+\Delta x_\tau$, but they are not equal: this is an unphysical result. Equation \eqref{eqn:spring initial stress to classical}, on the other hand, gives $U_\tau(\Delta x,F_\tau(\Delta x_\tau,0))=U_\tau(\Delta_x+\Delta x_\tau,0)$ and, in fact, more generally,
\begin{equation}\label{eqn:spring ISRI}
U_\tau(\Delta x,F_\tau(\Delta x_\tau,\tau))=U_\tau(\Delta x+\Delta x_\tau,\tau),
\end{equation}
for any value of $\tau$; therefore, this form of the stored energy function $U_\tau(\Delta x, \tau)$ produces consistent predictions of the amount of energy stored regardless of which reference configuration is chosen.

The preceding analysis clearly holds for initial stresses that arise from elastic deformations, but, in fact, the same arguments also apply to inelastic initial stresses. Consider the case where the spring has undergone an inelastic transformation such that it is no longer at its {original} natural length $x_0$ at equilibrium. It could be that it has been stretched or compressed, e.g. plastically or thermally, so that its {new natural} length is $L \ne x_0$ in the absence of external loads. We call $\tau$ the built-in (residual) force {when the spring is at its original length $x=x_0$}.

We will assume that Hooke's Law still applies to the stressed spring. Hence, there is a linear relationship between a force $F$ applied to the spring and its length $x$: $F = {a} x + {b}$, where $a$ and $b$ are constants. When $x={x_0}$, we have $F=\tau$ so that ${b}=\tau - {ax_0}$ and 
\begin{equation} \label{F-spring}
    F = {a}(x-{x_0}) + \tau.
\end{equation}

Assume that the potential energy is of the form $U_{x,\tau}(x,\tau)$. 
Now, do we have full latitude to choose any form of $U_\tau$ that leads to equation \eqref{F-spring}?
For example, both 
\begin{equation}
    U_{x,\tau} = \tfrac{1}{2}{a}(x-{x_0})^2 + \tau(x-{x_0}) + \frac{\tau^2}{2{a}}
    \qquad \text{and} \qquad
    \hat{U}_{x,\tau} = \tfrac{1}{2}{a}(x - {x_0})^2 + \tau(x - {x_0})
\end{equation}
lead to the correct expression for $F$.
However, the {latter} is not acceptable. To see this, we introduce {the new natural length} ${L}= {x_0} - \tau/{a}$, and rewrite \eqref{F-spring} as $F = {a} \Delta x$ where $\Delta x = x - {L}$ is the extension {relative to $L$}. Then, the same reasoning as earlier can be followed through, to the same conclusion, because the expressions for $U_{x,\tau}$ and $\hat{U}_{x,\tau}$, above, correspond to equations \eqref{eqn:spring initial stress to classical} and \eqref{eqn:spring guess}, respectively{, but with a different spring constant}.

The conclusion here is that we naturally reach a restriction that needs to be satisfied by the potential energy function $U_\tau = U_\tau(\Delta x, \tau)$. This will the theme of the following sections, but in three spatial dimensions. 


\section{Stored energy functions for initially strained materials}
\label{sec:strained energies}


We begin by investigating stored energy functions of the form $W := W(\mathbf{F}_2, \mathbf{F}_1, \vec{X}_2)$, where $\mathbf F_2$ is the elastic deformation gradient from the reference configuration $\bb_2$, $\mathbf F_1$ is the initial deformation gradient from the configuration $\bb_1$ to the configuration $\bb_2$, and $\vec X_2 \in \bb_2$ is the position in space, see \Cref{fig:Configurations}. We note that when the material is in the configuration $\bb_1$ or $\bb_2$ it may not be stress-free.


\subsection{Classical nonlinear elasticity}
\label{sec:classical}


First, we summarise one classical way~\cite{marsden1994mathematical} to deduce the functional forms of elastic stored energies $W$. 
We restrict our attention to isothermal deformations, and thus neglect the influence of temperature on the potential energy for simplicity, as it does not change any of the results presented here.

Let $\psi(\vec x)$ be the stored energy density in the current configuration, $\mathcal C$, per unit volume of $\mathcal C$. Let $\phi: \vec X \mapsto \vec x$ be a map describing the elastic deformation from the reference configuration $\mathcal R$ to $\mathcal C$. We remark  that the reference configuration $\mathcal R$ does not need to be stress-free. A set of classical assumptions \cite[Chapter 3 - Constitutive Theory]{marsden1994mathematical} implies that there exists a function $\tilde W$ in the form
\begin{equation} \label{eqn:classical_W}
{J}^{-1}\tilde W(\mathbf F, {\vec X})   = \psi(\vec x),
\end{equation}
for every $\mathbf F$, ${\vec X}$, $\vec{x}$ and $\phi$ such that  ${\mathbf F}=\partial\phi({\vec X}) / \partial {\vec X} =\partial{\vec x}/ \partial {\vec X}$, with $J = \det \mathbf F$. Throughout this paper we used the convention that quantities indicated with superscript tildes have the same meaning as the corresponding ones in \textit{classical} elasticity. Some of the basic assumptions used in the following include that the map $\phi$ is at least $C^2$ differentiable, and that the deformation $\phi$ conserves energy. 

The classical assumptions \cite{marsden1994mathematical} also imply that the Cauchy stress tensor in $\mathcal{C}$ is given by
\begin{equation}\label{eqn:classical_sigma}
	\tilde{\vec \sigma}(\mathbf{F},\vec{X}) = J^{-1} \frac{\partial \tilde W(\mathbf F, {\vec X})}{\partial \mathbf F} \mathbf F^{\mathrm T},
\end{equation}
where differentiation with respect to a tensor is defined component-wise as $(\partial/\partial\mathbf{A})_{ij}=\partial/\partial A_{ij}$, for any tensor $\mathbf{A}$ that has components $A_{ij}$ with respect to a given basis. 

\textbf{Locality assumption.} One assumption we want to draw attention to is \emph{locality}, see \cite[2.2 Axiom of Locality]{marsden1994mathematical}. It states that the energy density $\psi(\vec x)$ depends on the map $\phi: \vec X \mapsto \vec x$ only through quantities that are defined in the neighbourhoods of $\vec X$ and $\vec x$. This assumption is essential to deduce that $\tilde W$ depends on the map $\phi$ through the deformation gradient $\mathbf F$. 

We note that relaxing the assumptions in \cite[Chapter 3 - Constitutive Theory]{marsden1994mathematical} leads to other formulations such as either implicit constitutive  theories~\cite{rajagopal2003implicit}, where the stress and strain can be related through an implicit equation, or first and second gradient theories~\cite{mindlin1968first}, where the strain energy can depend on the gradient of the strain. Here we consider only the classical assumptions described in \cite{marsden1994mathematical}.


\subsection{A restriction on stored energy functions for initially strained materials}
\label{sec:restrict}

\begin{figure}[ht!]
\centering
\input{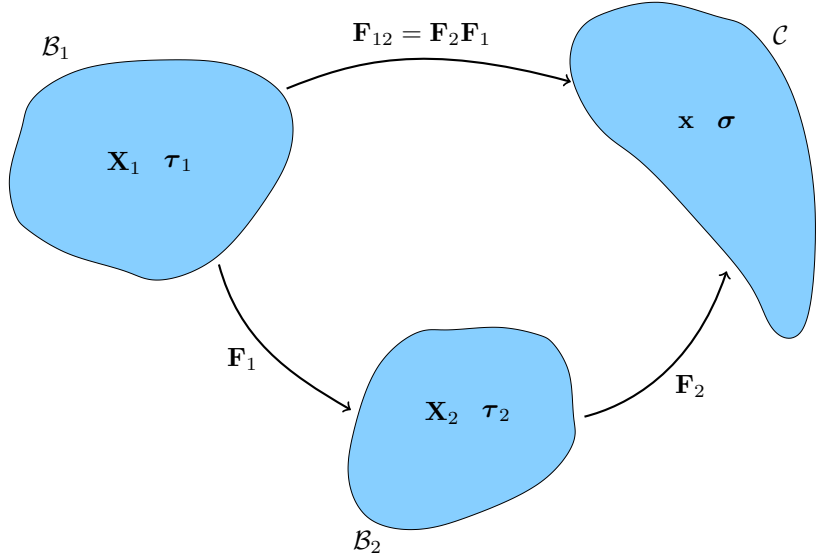}
\caption{One material can be elastically deformed between different configurations, $\bb_1$, $\bb_2$ and ${\mathcal C}$. We relate the configurations through elastic maps $\phi_1$, $\phi_2$ and $\phi_{12}$, where $\vec X_2 = \phi_1({\vec X}_1)$, and $\vec x = \phi_2({\vec X_2})=\phi_{12}(\vec{X}_1)$.  The deformation gradients are defined as $\mathbf F_1 = \partial \phi_1(\vec X_1)/\partial \vec X_1$, $\mathbf F_2 = \partial \phi_2(\vec X_2)/\partial \vec X_2$, and $\mathbf F_{12} = \partial \phi_{12}(\vec X_1)/\partial \vec X_1$ and their determinants are $J_1=\det\mathbf{F}_1$, $J_2=\det\mathbf{F}_2$ and $J_{12}=\det\mathbf{F}_{12}=\det(\mathbf{F}_1\mathbf{F}_2)=\det\mathbf{F}_1\det\mathbf{F}_2=J_1J_2$, respectively. The Cauchy stress tensors in $\bb_1$ and $\bb_2$ are $\vec \tau_1$ and $\vec \tau_2$, respectively.}
\label{fig:Configurations}
\end{figure}

We start by considering stored energy functions of the form $W(\mathbf F_2,\mathbf F_1, \vec X_2)$, where $\mathbf F_2$ is the elastic deformation gradient associated with deformations \textit{away from} the reference configuration being used, $\mathbf F_1$ is an initial deformation gradient associated with an elastic deformation \textit{into} the reference configuration, and $\vec X_2$ is a position vector in the reference configuration. See \Cref{fig:Configurations} for an illustration of these configurations.

We want $W$ to give the {stored} energy density, per unit volume of its reference configuration. That is, using $\bb_2$ as the reference configuration, see \Cref{fig:Configurations}, we want
\begin{equation} \label{def:W want}
	J_2^{-1}W(\mathbf F_2, \mathbf F_1, \vec X_2) = \psi(\vec x),
\end{equation}
where we recall that $\psi$ is the potential energy stored in the current configuration $\mathcal{C}$. Next, we show that the assumptions of classical elasticity imply that there does exist such as function $W$, where the above holds for every choice of $\mathbf F_1$, $\mathbf F_2$, and $\mathbf X_2$, 

Taking $\bb_1$ as the reference configuration, for \textit{any} elastic map $\phi_{12}$ and associated deformation gradient $\mathbf{F}_{12}$, the assumptions of classical elasticity imply that there exists a function $\tilde W$ such that
\begin{equation} \label{def:tilde W}
	J_{12}^{-1}\tilde W(\mathbf F_{12}, \vec X_1)=J_1^{-1}J_2^{-1}\tilde W(\mathbf F_2\mathbf F_1, \vec X_1) = \psi(\vec x),
\end{equation}
where $\tilde W$ is the stored energy per unit volume of $\bb_1$ (note that $\bb_1$ could be under stress). We emphasise that the above equation holds true for every $\mathbf F_1$ and $\mathbf F_2$, and that the function $\tilde{W}$ is strictly associated with the reference configuration $\bb_1$. 

By comparing~\eqref{def:W want} and \eqref{def:tilde W}, we conclude that as \eqref{def:tilde W} holds for every $\mathbf F_1$ and $\mathbf F_2$ then we can define:
\begin{equation} \label{def:W strain}
	W(\mathbf F_2, \mathbf F_1, \vec X_2) := J_1^{-1} \tilde W(\mathbf F_2 \mathbf F_1, \phi_1^{-1}(\vec X_2)) ,
\end{equation}
for every $\mathbf F_2$, $\mathbf F_1$, and $\vec X_2$. The above uniquely defines $W$, given $\tilde W$, because $\phi_1$ is invertible. In conclusion, 
\eqref{def:W want} holds for every choice for every $\mathbf F_2$, $\mathbf F_1$, and $\vec X_2$.

Equation~\eqref{def:W strain} naturally restricts the form of the function $W$, as follows. First note that equation \eqref{def:W strain} holds for \textit{any} values of the arguments of $W$, so we can set the first argument to be $\mathbf{F}_{12}$ (\i.e. $\mathbf{F}_2\rightarrow\mathbf{F}_{12}$) and the second argument to be the identity tensor $\mathbf{I}$ (\i.e. $\mathbf{F}_1\rightarrow\mathbf{I}$). Hence we are considering the initial map to be the trivial case of no initial deformation, which implies that $\vec X_2\rightarrow\vec X_1$, and, on the right side $J_1^{-1}\rightarrow1$, $\mathbf{F}_2\mathbf{F}_1\rightarrow\mathbf{F}_{12}\mathbf{I}$ and $\phi_1^{-1}(\vec{X}_2)\rightarrow\vec{X}_1$. Using these substitutions in \eqref{def:W strain}  leads to
\begin{equation}
    W(\mathbf F_{12}, \mathbf I, \vec X_1)=W(\mathbf F_2 \mathbf F_1, \mathbf I, \vec X_1)=
    \tilde W(\mathbf F_{12}\mathbf I,\vec X_1) = 
    \tilde W(\mathbf F_2 \mathbf F_1, \vec X_1).
\end{equation}
From \eqref{def:W strain}, we have that $\tilde W(\mathbf F_2 \mathbf F_1, \vec X_1) = J_1 W(\mathbf F_2, \mathbf F_1, \vec X_2)$, which used in the above leads to the restriction
\begin{equation} \label{eqn:strain_ISRI}
	\tcboxmath{
W(\mathbf F_2 \mathbf F_1, \mathbf I, \vec X_1)=J_1 W(\mathbf F_2, \mathbf F_1, \vec X_2),
}
\end{equation}
which must hold for every $\mathbf F_1$, $\mathbf F_2$, $\vec X_1$ and $\vec X_2$. Let us call this restriction \emph{Initial Strain Reference Independence}. See Appendix \ref{sec:alt} for an alternative derivation of this restriction. 
It can be shown that equation \eqref{eqn:spring initial stress to classical} satisfies the one-dimensional version of this restriction, whereas equation \eqref{eqn:spring guess} does not.

Next, we use equation \eqref{eqn:strain_ISRI} to derive a restriction on the Cauchy stress.
From~\eqref{def:W strain}, we can rewrite the Cauchy stress functional~\eqref{eqn:classical_sigma}, now as a function of three variables, in the form
\begin{equation} \label{eqn:stress initial strain}
	\vec \sigma(\mathbf F_2, \mathbf F_1,\vec{X}_2) := J_2^{-1} \frac{\partial W(\mathbf F_2, \mathbf F_1, \vec{X}_2)}{\partial \mathbf F_2} \mathbf F_2^{\mathrm T} = \tilde{\vec\sigma}(\mathbf{F}_{12},\vec{X}_1).
\end{equation}
To prove the latter equality, we use the chain rule as follows:
\begin{equation} \label{eqn:stress proof}
	J_2^{-1} \frac{\partial W(\mathbf F_2, \mathbf F_1, \mathbf X_2)}{\partial \mathbf F_2} \mathbf F_2^{\mathrm T} = J_2^{-1}\Big [
		J_1^{-1} \frac{\partial \tilde W(\mathbf F_{12}, \mathbf X_1)}{\partial \mathbf F_{12}}: \frac{\partial \mathbf F_{12}}{\partial \mathbf F_{2}}
	\Big]
	\mathbf F_2^{\mathrm T}=
	J_{12}^{-1}\frac{\partial \tilde W(\mathbf F_{12}, \vec X_1)}{\partial \mathbf F_{12}} \mathbf F_{12}^{\mathrm T} = \tilde{\vec\sigma}(\mathbf{F}_{12},\vec{X}_1),
\end{equation}
where the double-contraction operator is defined such that $(\mathbf{A}:\mathbf{B})_{kl}=A_{ij}B_{ijkl}$, for second- and fourth-order tensors $\mathbf{A}$ and $\mathbf{B}$, with components $A_{ij}$ and $B_{ijkl}$, respectively.

By substituting equations~\eqref{eqn:strain_ISRI} and ~\eqref{eqn:strain_ISRI2} into equation~\eqref{eqn:stress initial strain} and making use of some of the results from \eqref{eqn:stress proof}, we obtain the following restriction on the Cauchy stress functional:
\begin{equation} \label{eqn:strain ISS}
	\tcboxmath{
	\vec \sigma(\mathbf F_2, \mathbf F_1,\vec{X}_2) = 
 \vec \sigma(\mathbf F_2 \mathbf F_1, \mathbf I,\vec X_1),}
\end{equation}
for every $\mathbf F_1$, $\mathbf F_2$, $\vec X_1$, and $\vec X_2$.

In this section we deduced that, for all materials which satisfy the classical assumptions of elasticity, as described in \Cref{sec:classical}, there exists a function $W$ that satisfies~\eqref{eqn:strain_ISRI} for every initial deformation gradient $\mathbf{F}_1$ and subsequent elastic deformation gradient $\mathbf{F}_2$.


\section{Stored energy functions for initially stressed materials}
\label{sec:initially stress energies}


In this section we move from {stored} energy functions of the form $W(\mathbf F_2, \mathbf F_1, \vec X_2)$ to stored energy functions of the form $\Wtau(\mathbf F_2, \vec \tau_2, \vec X_2)$, where $\vec\tau_2$ is the initial stress in $\bb_2$, see Figure \ref{fig:Configurations}. We show that the Initial Strain Reference Independence restriction on $W$ can be used to derive a restriction on $\Wtau$ by connecting $\vec\tau_2$ to an initial stretch tensor $\mathbf U_1$, corresponding to a deformation from a (potentially virtual) configuration $\bb_1$ to $\bb_2$. We find that, for an isotropic material, this restriction is equivalent to that introduced in~\cite{gower2017new}, which we called Initial \textit{Stress} Reference Independence (ISRI).


\subsection{Initial stress from initial strain}
\label{sec:stress energy from strain}


Using the polar decomposition theorem~\cite{higham1990fast}, we can write $\mathbf F_1$ uniquely as $\mathbf F_1 = \mathbf R_1 \mathbf U_1$, where $\mathbf U_1 = \mathbf U_1^{\mathrm T}$ is the right stretch tensor and $\mathbf R_1$ is a proper rotation tensor. Combining the polar decomposition with~\eqref{eqn:classical_sigma}, it can be shown \cite{ogden1997non} that 
\begin{equation} \label{eqn:sigma U}
	\tilde{\vec \sigma}(\mathbf F_1, \vec X_1) =  \mathbf R_1 \tilde{\vec \sigma}(\mathbf U_1, \vec X_1) \mathbf R_1^{\mathrm T},
\quad \text{where} \;\;\
\tilde{\vec \sigma}(\mathbf U_1, \vec X_1) := J_1^{-1} \frac{\partial \tilde W(\mathbf U_1, {\vec X_1})}{\partial \mathbf U_1} \mathbf U_1,
\end{equation}
by using the fact that $\tilde W(\mathbf R_1 \mathbf U_1, {\vec X_1}) = \tilde W(\mathbf U_1, {\vec X_1})$ due to objectivity.

To define an appropriate form for the stored energy function, $\Wtau(\mathbf F_2, \vec \tau_2,\vec X_2)$, we need, for any stress tensor $\vec\tau_2$ and position vector $\vec X_2$ in $\bb_2$, a function $\tilde {\vec \sigma}^{-1} (\vec \tau_2,\vec X_2) = \mathbf U_1$ that gives the stretch tensor $\mathbf{U}_1$ that led to $\vec\tau_2$ from $\bb_1$. In \Cref{app:inverse stress}, we explain why an inverse function $\tilde {\vec \sigma}^{-1}$ should exist, and can be uniquely defined, for a wide range of constitutive choices (those that are stable under traction). However, to simplify the discussion below, we simply \textit{assume} that such an inverse function exists. An example of an inverse function $\tilde {\vec \sigma}^{-1}$ is given for third-order energies in \Cref{sec:initial stress third-order} by equation \eqref{eqn:strain to stress}.

As another example for which an inverse function does exist, consider the stress-stretch relation for an incompressible neo-Hookean material: $\vec \tau_2= -p\vec I + \mu \vec F_1\vec F_1^\text{T} = \vec R_1\tilde {\vec \sigma}(\vec U_1) \vec R_1^\text{T}$, where $\tilde {\vec\sigma}  = - p\vec I + \mu \vec U_1^2$ ($p$ is a Lagrange multiplier and $\mu$ is the shear modulus). This equation can be  inverted to find $\vec U_1 = \tilde {\vec \sigma}^{-1} (\vec \tau_2, \vec X_2) = [(\vec \tau_2 + p_0\vec I)/\mu]^{1/2}$, where $p_0$ is expressed in terms of $\vec \tau_2$ (see \cite[Section 3]{gower2015initial} for details). A similar inversion to find $\vec U_1^2$ in terms of $\vec \tau_2$ can be achieved for the Mooney-Rivlin model \cite{agosti2018constitutive,johnson1993dependence} and, indeed, for any hyperelastic, isotropic model \cite{johnson1993dependence}, so that $\vec U_1$, the square root of $\vec U_1^2$, can be expressed explicitly \cite{hoger1984determination}.

Using the inverse function, $\tilde {\vec \sigma}^{-1}$, we can define a stored energy function, $\Wtau$, that depends on strain and initial stress, as
\begin{equation} \label{def:W tau}
	W_\tau(\mathbf F_2, \vec \tau_2, \vec X_2) := W(\mathbf F_2, \tilde {\vec \sigma}^{-1}(\vec \tau_2,\vec X_2) , \vec X_2).
\end{equation}
We then define a new functional for the Cauchy stress, which now depends on the initial \textit{stress} $\vec\tau_2$ instead of the initial \textit{strain} $\mathbf{F}_1$, as follows
\begin{equation} \label{eqn:stress initial stress}
\sigmatau(\mathbf F_2, \vec \tau_2,\vec X_2) := J_2^{-1} \frac{\partial \Wtau(\mathbf F_2, \vec \tau_2, \vec X_2)}{\partial \mathbf F_2} \mathbf F_2^{\mathrm T} = \vec \sigma(\mathbf F_2, {\vec \sigma}^{-1}(\vec \tau_2,\vec X_2),\vec X_2),
\end{equation}
where $\vec\sigma(\mathbf{F}_2,{\vec \sigma}^{-1}(\vec \tau_2,\vec X_2),\vec X_2)$ is defined in equation~\eqref{eqn:stress initial strain};  the second equality follows from equation~\eqref{def:W tau}.

We are now ready to prove a restriction on $\Wtau$. In equation \eqref{eqn:stress initial stress}, which is valid for any values of its arguments, we let $\mathbf{F}_2\rightarrow\mathbf{U}_1$, $\vec\tau_2\rightarrow\vec\tau_1$ and $\vec X_2\rightarrow\vec X_1$ to obtain
\begin{equation}
\sigmatau(\mathbf U_1, \vec \tau_1,\vec X_1) = \vec \sigma(\mathbf U_1, \tilde{\vec\sigma}^{-1}(\vec \tau_1,\vec X_1),\vec X_1) = \vec \sigma(\mathbf U_1, \mathbf{I},\vec X_1) = \tilde{\vec \sigma}(\mathbf U_1,\vec X_1),
\end{equation}
where the latter equality follows from equation~\eqref{eqn:stress initial strain}. We substitute this equation into equation~\eqref{def:W tau} to obtain
\begin{equation} \label{eqns:ISRI_step_1}
J_1 \Wtau(\mathbf F_2, {\sigmatau}(\mathbf U_1, \vec \tau_1,\vec X_1), \vec X_2)
 = J_1 \Wtau(\mathbf F_2, \tilde{\vec \sigma}(\mathbf U_1,\vec X_1), \vec X_2)
 = J_1 W(\mathbf F_2, \mathbf U_1, \vec X_2).
\end{equation}
We can rewrite the last term above by using \eqref{eqn:strain_ISRI} followed by~\eqref{def:W tau} to obtain
\begin{equation}  \label{eqns:ISRI_step_2}
J_1 W(\mathbf F_2, \mathbf U_1, \vec X_2)
= W(\mathbf F_2\mathbf U_1,\mathbf I,{\vec X_1})
= \Wtau(\mathbf F_2\mathbf U_1, \vec \tau_1, {\vec X_1}).
\end{equation}

Finally, equating the last term of  \eqref{eqns:ISRI_step_2} with the first term in \eqref{eqns:ISRI_step_1} leads to
\begin{gather}\label{eqn:stress_ISRI}
	\tcboxmath{
 \Wtau(\mathbf F_2\mathbf U_1, \vec \tau_1, \vec X_1)=J_1 \Wtau(\mathbf F_2, {\sigmatau}(\mathbf U_1, \vec \tau_1,\vec X_1),  \vec X_2),
	\quad \text{(ISRI restriction)}
	}
\end{gather}
which holds for every $\mathbf F_2$, $\mathbf U_1$ and $\vec\tau_1$. This restriction was first derived in a slightly different form in \cite{gower2017new}. The version in \cite{gower2017new} assumed that there was no structural anisotropy in the material, which allows this equation to be rewritten in terms of $\mathbf{F}_1$ instead of $\mathbf U_1$, as we show in the next section. 

{We note that the restriction~\eqref{eqn:stress_ISRI} holds for every value of the \textit{three} variables $\mathbf F_2$, $\mathbf U_1$, and $\vec \tau_1$. However, we only wish to restrict the dependence of $\Wtau$ on its first \textit{two} arguments. Therefore, the restriction~\eqref{eqn:stress_ISRI} has a redundant degree of freedom. As a consequence, it is sufficient to enforce the restriction \eqref{eqn:stress_ISRI} for every $\mathbf F_2$ and $\mathbf U_1$, while $\vec \tau_1$ remains fixed. For example, if we take $\vec \tau_1 = \vec 0$, we obtain
\begin{gather}\label{eqn:stress_ISRI zero tau}
	 \Wtau(\mathbf F_2\mathbf U_1, \vec 0, \vec X_1) = J_1 \Wtau(\mathbf F_2, {\sigmatau}(\mathbf U_1, \vec 0,\vec X_1),  \vec X_2),
\end{gather}
for every $\mathbf F_2$, $\mathbf U_1$. Here, it is clear that the left side is equivalent to a classical strain energy function, and therefore \eqref{eqn:stress_ISRI zero tau} enforces $\Wtau(\mathbf F_2, \vec \tau_2, \vec X_2)$ to be equivalent to a classical strain energy function. 
}

Finally, differentiating both sides of \eqref{eqn:stress_ISRI} with respect to $\mathbf F_2$, and using~\eqref{eqn:stress initial stress} and \eqref{eqn:stress initial strain}, we obtain
\begin{gather}\label{eqn:stressU_ISS}
	\tcboxmath{
\sigmatau(\mathbf F_2 \mathbf U_1, \vec \tau_1,\vec X_1)=	\sigmatau(\mathbf F_2, {\sigmatau}(\mathbf U_1, \vec \tau_1,\vec X_1),\vec X_2),
	\quad \text{(Initial Stress Symmetry)}
	}
\end{gather}
which was first introduced (again in a slightly different form) in~\cite{gower2015initial}.


\subsection{Material symmetry}


We now specialise the results above to the case where there is no underlying \textit{structural} anisotropy in the material (so that the material would be isotropic in the absence of initial stress). In this case, the value of the potential energy does not change after a rigid-body rotation of the reference configuration of $\mathbf F$ which leads to
\begin{equation}\label{eqn:initial isotropy}
\Wtau(\mathbf F, \vec \tau,\vec X) = \Wtau(\mathbf F \mathbf R^\mathrm{T}, \mathbf R \vec \tau \mathbf R^\mathrm{T},\vec X),
\end{equation}
for every $\mathbf{F}$, $\vec\tau$, $\vec X$ and orthogonal $\mathbf R$. Using this result in equation~\eqref{eqn:stress initial stress}, we also have that
\begin{equation}\label{eqn:stress_objectivity}
\sigmatau(\mathbf F, \vec \tau,\vec X) = \sigmatau(\mathbf F \mathbf R^\mathrm{T}, \mathbf R \vec \tau \mathbf R^\mathrm{T},\vec X),
\end{equation}
for every $\mathbf{F}$, $\vec\tau$, $\vec X$ and orthogonal $\mathbf R$. By defining $ \vec \tau_R  = \mathbf R_1^\mathrm{T} \vec \tau_1 \mathbf R_1$, where $\mathbf{R}_1$ is an orthogonal tensor that satisfies $\mathbf{F}_1=\mathbf{U}_1\mathbf{R}_1^\mathrm{T}$, and using equations \eqref{eqn:initial isotropy} and \eqref{eqn:stress_objectivity} we obtain:
\begin{align}
	\Wtau(\mathbf F_2 \mathbf U_1, \vec \tau_R, \vec X_1) & = \Wtau(\mathbf F_2 \mathbf U_1 \mathbf R_1^\mathrm{T}, \mathbf R_1 \vec\tau_R \mathbf R_1^\mathrm{T},\vec X_1),
 \\	J_1 \Wtau(\mathbf F_2, {\sigmatau}(\mathbf U_1, \vec \tau_R,\vec X_1), \vec X_2) & =
	J_1 \Wtau(\mathbf F_2, {\sigmatau}(\mathbf U_1 \mathbf R_1^\mathrm{T}, \mathbf R_1 \vec \tau_R \mathbf R_1^\mathrm{T},\vec X_1),\vec X_2).
\end{align}
The terms on the left side of both equations are equal due to ISRI~\eqref{eqn:stress_ISRI}; therefore, the terms on the right side are also equal. Equating the terms on the right, and using $\mathbf F_1  = \mathbf U_1 \mathbf R_1^\mathrm{T}$ and $\mathbf R_1 \vec\tau_R \mathbf R_1^\mathrm{T}=\vec\tau_1$ leads to
\begin{equation} \label{eqn:stress_ISRI_isotropy}
\tcboxmath{
 \Wtau(\mathbf F_2 \mathbf F_1,\vec \tau_1, \vec X_1)=	J_1 \Wtau(\mathbf F_2, {\sigmatau}(\mathbf F_1 , \vec \tau_1,\vec X_1),\vec X_2),
	}
\end{equation}
for every $\mathbf F_2$, $\mathbf F_1$ and $\vec \tau_1$. This form of ISRI is the same as that which first appeared~\cite{gower2017new}, now including explicit dependence on the position vectors. Similarly, in this case, equation \eqref{eqn:stressU_ISS} can be written as
\begin{gather}\label{eqn:stressF_ISS}
	\tcboxmath{
\sigmatau(\mathbf F_2 \mathbf F_1, \vec \tau_1,\vec X_1)=	\sigmatau(\mathbf F_2, {\sigmatau}(\mathbf F_1, \vec \tau_1,\vec X_1),\vec X_2).}
\end{gather}

See \cite{rajagopal2024residual} for some comments on material symmetry for constitutive equations with residual stress. However, we take the view that the constitutive choice should hold for all forms of initial stress, and not just 
 for residual stress. 


\section{Third-order stored energy functions}
\label{sec:third-order energies}


In this section, we discuss stored energy functions that depend on strain and stress up to third order in an expansion for small values of these arguments. We show that these stored energy functions are only consistent with classical third-order strain energy functions when {imposing either the restriction~\eqref{eqn:strain_ISRI} or the restriction~\eqref{eqn:stress_ISRI_isotropy}}.  

In classical third-order elasticity theory, the strain energy function $\tilde{W}$ is expanded in powers of the strain. For example, if we consider the deformation from $\bb_1$ to $\mathcal{C}$ in Figure \ref{fig:Configurations}, this time, assuming the deformation is infinitesimal, then the strain energy function can be written as  
\begin{equation} \label{eqn:classical-third}
\tilde W_\mathbf{E}(\mathbf E_{12}) = \frac{\lambda}{2}\left(\tr \,\Eh_{12}\right)^2 + \mu\,\tr\,\Eh_{12}^2 + \frac{A}{3}\tr \, \Eh_{12}^3 + B\,\tr\,\Eh_{12}\,\tr\,\Eh_{12}^2 + \frac{C}{3}(\tr \,\Eh_{12})^3,
\end{equation}
 where $\lambda, \mu$ are the Lam\'e parameters, $A, B, C$ are the Landau parameters, and $\Eh_{12} = \tfrac{1}{2} \left(\mathbf F_{12}^\text{T} \mathbf F_{12} - \mathbf I\right)= \tfrac{1}{2} \left(\mathbf F_1^\text{T} \mathbf F_2^\text{T} {\mathbf F_2} {\mathbf F_1} - \mathbf I\right)$ is the Green-Lagrange strain tensor. Note that there are other, equivalent, expansions \cite{wikiconstants}, all involving five independent elastic constants. In this section, we assume that $\Eh_{12}$ is small, so that the expansion \eqref{eqn:classical-third} is valid up to $\mathcal O(\mathbf E_{12}^3)$. We also assume that the material is homogeneous in $\bb_1$, so we can omit the dependence on the position vectors, and that $\bb_1$ is stress-free, so that $\vec \tau_1 = \vec 0$. The strain energy function \eqref{eqn:classical-third} is deduced by assuming that $\bb_1$ is stress-free, so we make this assumption throughout this section for consistency; however, we reiterate that the restrictions \eqref{eqn:strain_ISRI} and \eqref{eqn:stress_ISRI} do not assume that $\bb_1$ is stress-free.

{
For third-order elasticity \cite{hughes1953second,murnaghan1937finite}, the most common way to understand how an initial stress $\vec \tau$ \textit{affects} an elastic deformation is to consider an initial stress that is \textit{due} to an elastic deformation. That is, to assume $\vec \tau = \tilde{\vec \sigma}(\mathbf F_{1})$, and then take a further elastic deformation $\mathbf F_2$ on top of this stressed state. Following this route, we can define the initially stressed stored energy function $\Wtau$ such that, for all values of $\mathbf F_2$ and $\vec \tau = \tilde{\vec \sigma}(\mathbf F_{1})$, it satisfies the following equation:
\begin{equation} \label{eqn:classic stressed third}
    J_1 \Wtau (\mathbf F_2, \vec \tau) = \tilde W_{\mathbf E}(\mathbf E_{12}),
\end{equation}
where $\tilde W_{\mathbf E}$ is given by equation \eqref{eqn:classical-third}. The previous sections of this paper have demonstrated that stored energy functions of the form $\Wtau (\mathbf F_2, \vec \tau)$ must satisfy the ISRI restriction~\eqref{eqn:stress_ISRI zero tau}. For clarity, we repeat those steps here in the case where $\mathbf F_1 = \mathbf I$, so that $\vec \tau = \vec 0$ and \eqref{eqn:classic stressed third} becomes $\Wtau (\mathbf F_2, \vec 0) = \tilde W(\mathbf E_{2})$ for every $\mathbf F_2$, where $\mathbf E_{2} = \frac{1}{2}(\mathbf F_2^\text{T} \mathbf F_2 - \mathbf I)$. In this case, $\mathbf F_2=\mathbf F_2 \mathbf F_1=\mathbf{F}_{12}$, which implies that $\Wtau (\mathbf F_2 \mathbf F_1, \vec 0)=\Wtau(\mathbf F_{12},\vec 0) = \tilde W(\mathbf E_{12})$, and, therefore, $J_1  \Wtau (\mathbf F_2, \vec \tau) = \Wtau(\mathbf F_{12},\vec 0)$, which is equivalent to the
restriction~\eqref{eqn:stress_ISRI_isotropy} in the case where $\mathbf{F}_1=\mathbf{I}$. That is, an initially stressed stored energy function defined via equation \eqref{eqn:classic stressed third} automatically satisfies the ISRI restriction.
}

{
Below, we show that, without the ISRI restrictions, if we take a naive expansion of the stored energy function in terms of its invariants, we can obtain third-order stored energy functions which are not equivalent to \eqref{eqn:classical-third} and have too many free parameters. The ISRI restrictions are necessary to resolve this inconsistency.

In the following, we undertake the asymptotic expansions in order of growing complexit,  starting first from an initially \textit{strained} third-order energy function and later dealing with an  initially \textit{stressed} third-order energy function.
}

\subsection{{An initially strained, third-order stored energy function}}

Here, we deduce a third-order stored energy function of the form $W(\mathbf{F}_2,\mathbf{F}_1)=\We(\mathbf E,\mathbf e)$, a function of the small initial strain $\mathbf E = \tfrac{1}{2}(\mathbf F_2^\text{T}\mathbf F_2 - \mathbf{I})$ and the small subsequent strain $\mathbf e = \tfrac{1}{2}(\mathbf F_1 \mathbf F_1^\text{T} - \mathbf I)$.

Let us expand $\We(\mathbf{F}_2,\mathbf{F}_1)$ up to third order. Since this stored energy density depends only on $\mathbf E$ and $\mathbf e$, we can write it in terms of the nine independent mixed invariants of $\mathbf E$ and $\mathbf e$~\cite{shams2011initial, shariff2017spectral}:
\begin{align}\label{eqns:strain stress invariants}
    & I_1 = \tr(\mathbf E), && I_2 =\tr(\mathbf E^2), && I_3= \tr(\mathbf E^3),
    \notag \\
    & i_4 =\tr(\mathbf e), && i_5 = \tr(\mathbf e^2), && i_6 = \tr(\mathbf e^3),
 \notag \\     
    & i_7 = \tr(\mathbf e\mathbf E), && i_8 = \tr(\mathbf e^2 \mathbf E), && i_9 = \tr(\mathbf e\mathbf E^2).
\end{align}

The Cauchy stress tensor arising from $\We(\mathbf E,\mathbf e)$ can be written as
\begin{align}
&    \vec\sigma(\mathbf F_2,\mathbf{F}_1)=J_2^{-1}\mathbf{F}_2\left(\frac{\partial \We}{\partial I_1}\mathbf{I}+2\frac{\partial \We}{\partial I_2}\mathbf{E}+3\frac{\partial \We}{\partial I_3}\mathbf{E}^2+\frac{\partial \We}{\partial i_7}\mathbf{e}+\frac{\partial \We}{\partial i_8}\mathbf{e}^2+\frac{\partial \We}{\partial i_9}(\mathbf{Ee}+\mathbf{eE})\right)\mathbf{F}_2^\text{T}.
\end{align}

{
The next step is to assume that both $\mathbf e$ and $\mathbf E$ are small and to expand $\We(\mathbf E,\mathbf e)$ asymptotically up to $\mathcal O(\mathbf E^3)$. Here, we choose $\mathbf e = \mathcal O(\mathbf E)$, but we note that if we assumed, for example, that $\mathbf e$ is asymptotically smaller (e.g. $\mathbf e = \mathcal O(\mathbf E^2)$), or larger (e.g. $\mathbf e = \mathcal O(\mathbf E^\frac{1}{2})$), than $\mathbf E$, then the calculations below would be different, but the main result would be unchanged, namely: without imposing ISRI, $\We(\mathbf E,\mathbf e)$ has too many free constants, whereas only five independent constants remain after imposing ISRI, consistently with classical third-order elasticity. 
}

Let $\mathbf e = O(\mathbf E)$, we now follow a systematic expansion method~\cite{destrade2010third} to expand $\We(\mathbf E,\mathbf e)$ up to $O(\mathbf E^3)$ to obtain
\begin{multline}\label{eqn:W e}
    \We(\mathbf E, \mathbf e) =	
    \alpha_0I_1+\alpha_{1} I_1^2 + \alpha_{2} I_2  +
    \alpha_{3}  I_3 + \alpha_{4} I_1 I_2  + \alpha_{5} I_1^3
    + \alpha_{6} i_4  +  \alpha_{7} i_4^2  + \alpha_{8} i_5
    + \alpha_{9} i_6  + \alpha_{10} i_4 i_5 + 
    \alpha_{11} i_4^3 
\\ 
    + \alpha_{12}  i_9 + \alpha_{13} I_1^2 i_4  + \alpha_{14} I_1 i_7  + \alpha_{15} I_2 i_4+ \alpha_{16}i_7+\alpha_{17}I_1i_4+\alpha_{18}I_1i_4^2+\alpha_{19}I_1i_5+\alpha_{20}i_4i_7+\alpha_{21}i_8,
\end{multline}
where $\alpha_j$, for $j=0, \ldots, 21)$, are constants. Note that the requirement $\vec\sigma(\mathbf{I},\mathbf{I}) = \mathbf{0}$ implies that $\alpha_0=0$. This leaves a total of 21 free constants, whereas the strain energy function of classical third-order elasticity \eqref{eqn:classical-third} has only five independent constants. These two stored energy functions are expanded to the same asymptotic order, and account
for the same quantities, yet have a difference of 16 degrees of freedom, which is clearly inconsistent. This simple observation highlights the \textit{need for a restriction} to be imposed on stored energy functions of the form $\We(\mathbf E,\mathbf e)$. This restriction is provided by~\eqref{eqn:strain_ISRI}. 

The restriction \eqref{eqn:strain_ISRI} can be written in terms of the quantities introduced in this section as
\begin{equation} \label{eqn:strain_ISRI_E}
	\tcboxmath{
\We(\mathbf E_{12}, \mathbf 0)=J_1 \We(\mathbf E, \mathbf e),
}
\end{equation}
for every $\mathbf E$ and $\mathbf e$. The left side is
\begin{equation}
    \We(\mathbf E_{12}, \mathbf 0)=\alpha_0\tr\mathbf E_{12}+\alpha_1(\tr\mathbf E_{12})^2+\alpha_2\tr\mathbf E_{12}^2+\alpha_3\tr\mathbf E_{12}^3+\alpha_4\tr\mathbf E_{12}\tr\mathbf E_{12}^2+\alpha_5(\tr\mathbf E_{12})^3,
\end{equation}
which we can rewrite in terms of the mixed invariants of $\mathbf E$ and $\mathbf e$. Assuming that $\mathbf{E}$ is small and that $\mathbf{e} = \mathcal O(\mathbf E)$ we find
\begin{align} \label{eqn:tilde trE} 
& \tr \, {\mathbf E_{12}} = \tr \,\mathbf E + \tr \,\mathbf e + 2 \, \tr \,(\mathbf e \mathbf E) = I_1+i_4+2I_7, 
\notag \\
& \tr \, (\mathbf E_{12}^2) \approx \tr \,(\Eh^2) + \tr \, (\mathbf e^2) + 2 \,\tr \, (\mathbf e \Eh) +  + 4 \,\tr \, (\mathbf e^2 \Eh ) + 4\, \tr \, (\mathbf e \Eh^2) = I_2 + i_5 + 2i_7 + 4i_8 + 4i_9,
\notag  \\
 & \tr \, \Eh_{12}^3 \approx \tr \, (\Eh^3) + \tr\, (\mathbf e^3) + 3 \, \tr \, (\mathbf e^2 \Eh) + 3 \,\tr \, (\mathbf e \Eh^2) = I_3 + i_6 + 3i_8 + 3i_9,
\end{align}
where the first equality is exact and the other two are correct up to the third order in the strain measures. 
We then expand $J_1$ as
\begin{equation} \label{eqn:J_1 expand}
J_1 = \det \mathbf F_1 = \sqrt{\det(\mathbf I + 2\mathbf e)} = 1 + \tr \, \mathbf e +\tfrac{1}{2}(\tr\mathbf e)^2-\tr\mathbf e^2+O(\mathbf e^3)=1+i_4+\tfrac{1}{2} i_4^2 -i_5+O(\mathbf e^3),
\end{equation}
substitute this expression into the right side of equation \eqref{eqn:strain_ISRI_E}, neglect $O(\mathbf E^4)$ terms, and compare the coefficients multiplying each linearly independent combination of the invariants. This results in the following set of equations:
\begin{align}
    & \alpha_6=0,\quad\alpha_7=\alpha_1,\quad\alpha_8=\alpha_2,\quad\alpha_9=\alpha_3,\quad\alpha_{10}=\alpha_4-\alpha_2,\quad\alpha_{11}=\alpha_5-\alpha_1,\quad\alpha_{12}=4\alpha_2+3\alpha_3,
\notag \\
    & \alpha_{13}=3\alpha_5-\alpha_1,\quad\alpha_{14}=2(2\alpha_1+\alpha_4),\quad\alpha_{15}=\alpha_4-\alpha_2,\quad\alpha_{16}=2\alpha_2,\quad\alpha_{17}=2\alpha_1,\quad \alpha_{18}=3\alpha_5 -2 \alpha_1,\notag
\\ \label{eqn:alphas}
    & \alpha_{19}=\alpha_4,\quad\alpha_{20}=2(2\alpha_1-\alpha_2+\alpha_4),\quad\alpha_{21}=4\alpha_2+3\alpha_3.
\end{align}
A Mathematica file is provided as supplementary material to verify these calculations. These 16 non-trivial equations reduce the number of free parameters down to five, which we can write in terms of classical third-order constants. To rewrite them in terms of the Lam\'e and Landau parameters we simply set $\tilde{W}(\mathbf{E}_{12})=\We(\mathbf{E}_{12},\mathbf{0})$ to find
\begin{equation}
    \alpha_1=\frac{\lambda}{2},\quad\alpha_2=\mu,\quad\alpha_3=\frac{A}{3},\quad\alpha_4=B,\quad\alpha_5=\frac{C}{3}.
\end{equation}

For completeness, with the parameters relabelled as above, the stored energy function is given by:
\begin{multline}\label{eqn:WEe}
    \We(\mathbf E,\mathbf e) = \frac{\lambda}{2}I_1^2 + \mu I_2 + \frac{A}{3}I_3 + BI_1I_2+\frac{C}{3}I_1^3 + \frac{\lambda}{2}(1 - i_4)i_4^2 + \frac{C}{3} i_4^3  + \mu i_5 + \frac{A}{3}i_6+(B-\mu)i_4i_5  +(4\mu+A)i_9+\\
    \left(C-\frac{\lambda}{2}\right)I_1^2i_4+2(\lambda+B)I_1i_7+(B-\mu)I_2i_4+2\mu i_7+\lambda I_1i_4+(C-\lambda)I_1i_4^2+BI_1i_5+2(\lambda-\mu+B)i_4i_7+(4\mu+A)i_8.
\end{multline}

This seciton proves that using an expansion of the stored energy function in terms of the invariants of $\mathbf{E}$ and $\mathbf{e}$, as shown in equation \eqref{eqn:W e}, leads to too many degrees of freedom. This is expected because, as shown in \Cref{sec:restrict}, stored energy functions of the form $\We(\mathbf E, \mathbf e)$ need to satisfy the restriction \eqref{eqn:strain_ISRI}. Thus, we showed that the restriction \eqref{eqn:strain_ISRI} naturally implies that all third-order expansions of $\We(\mathbf E, \mathbf e)$ are equivalent to the classical third-order elastic strain energy function. 

Next, we consider third-order expansions of stored energy functions of the form {$\Wtau(\mathbf{F}_2,\vec \tau)=\hat{W}_{\vec\tau}(\mathbf E,\vec \tau)$}. Again, we show that a restriction is needed in the form of \eqref{eqn:stress_ISRI_isotropy} to reduce the number of free parameters and make this initially stressed stored energy function equivalent to a classical third-order strain energy function.

\subsection{{An initially stressed, third-order stored energy function}} 
\label{sec:initial stress third-order}

Let us consider a third-order expansion of a stored energy function of the form {$\hat{W}_{\vec\tau}(\mathbf E,\vec \tau)$} by assuming that both the stress $\vec \tau$ and the strain $\mathbf E = \tfrac{1}{2}(\mathbf F_2^\text{T}\mathbf F_2 - \mathbf{I})$ are small. To make this assumption rigorous, the stored energy function and stress need to be made dimensionless, for example by dividing them with respect to $\mu$, but we omit the details of this process here for brevity. We assume that the stored energy density depends on only $\mathbf E$ and $\vec \tau$, which implies that we can write $\hat{W}_{\vec\tau}(\mathbf E,\vec \tau)$ in terms of the nine independent invariants of $\mathbf E$ and $\vec \tau$~\cite{shams2011initial, shariff2017spectral}:
\begin{align}
    & I_1 = \tr(\mathbf E), && I_2 =\tr(\mathbf E^2),&& I_3= \tr(\mathbf E^3),
\notag\\ 
    & I_4 =\tr(\vec \tau), && I_5 = \tr(\vec \tau^2), && I_6 = \tr(\vec \tau^3), 
\notag\\
    & I_7 = \tr(\vec \tau  \mathbf E), && I_8 = \tr(\vec \tau^2 \mathbf E), && I_9 = \tr(\vec \tau \mathbf E^2).
\end{align}

The Cauchy stress arising from $\hat{W}_{\vec\tau}(\mathbf E,\vec \tau)$ can be expressed in terms of these invariants, as
\begin{align}
    & \sigmatauhat(\mathbf{F}_2,\vec\tau)=J_2^{-1}\mathbf{F}_2\left(\frac{\partial \hat{W}_{\vec\tau}}{\partial I_1}\mathbf{I}+2\frac{\partial \hat{W}_{\vec\tau}}{\partial I_2}\mathbf{E}+3\frac{\partial \hat{W}_{\vec\tau}}{\partial I_3}\mathbf{E}^2+\frac{\partial \hat{W}_{\vec\tau}}{\partial I_7}\vec\tau+\frac{\partial \hat{W}_{\vec\tau}}{\partial I_8}\vec\tau^2+\frac{\partial \hat{W}_{\vec\tau}}{\partial I_9}(\mathbf{E}\vec\tau+\vec\tau\mathbf{E})\right)\mathbf{F}_2^\text{T}.
\end{align}

We now assume that $\vec \tau = O(\mathbf E)$ and expand  $\hat{W}_{\vec\tau}(\mathbf E,\vec \tau)$ up to $O(\mathbf E^3)$. As discussed in the previous section, there are other choices other than $\vec \tau = O(\mathbf E)$, but the main message would be the same. Again, we follow a systematic expansion method similar to that given in~\cite{destrade2010third} to obtain
\begin{multline}\label{eqn:W tau}
\hat{W}_{\vec\tau}(\mathbf E, \vec \tau) =\beta_0{I_1}+
\beta_{1} I_1^2 + \beta_{2} I_2  +
\beta_{3}  I_3 + \beta_{4} I_1 I_2  + \beta_{5} I_1^3 
+ \beta_{6} I_4  +  \beta_{7} I_4^2  + \beta_{8} I_5
+ \beta_{9} I_6  + \beta_{10} I_4 I_5 + \beta_{11} I_4^3
\\
+ \beta_{12}  I_9 + \beta_{13} I_1^2 I_4  + \beta_{14} I_1 I_7  + \beta_{15} I_2 I_4+\beta_{16}I_7+\beta_{17}I_1I_4+\beta_{18}I_1I_4^2+\beta_{19}I_1I_5+\beta_{20}I_4I_7+\beta_{21}I_8,
\end{multline}
where $\beta_j$, for $j=0, \ldots, 21$, are constants.  A useful feature of stored energy functions of the above form is that they lead to simple expressions for elastic wave speeds in terms of the initial stress, which facilitates measuring the initial stress \cite{li2020ultrasonic}.

Before applying the ISRI restriction \eqref{eqn:stress_ISRI}, we can first reduce the number of free coefficients by enforcing stress compatibility: $\sigmatauhat(\mathbf{I},\vec\tau)=\vec\tau$ for every $\vec \tau$, which implies that $\beta_{16}=1$ and $\beta_{0} = \beta_{17}=\beta_{18}=\beta_{19}=\beta_{20}=\beta_{21}=0$. See Shams \textit{et al.}~\cite{shams2011initial} for further details on initial stress compatibility. 
 
We see that there are still 15 constants left, whereas there are only five constants for classical third-order elasticity \eqref{eqn:classical-third}. Again, this is unexpected and illustrates that a restriction is needed to reduce the number of constants in the stored energy function~\eqref{eqn:W tau}. Since we have assumed that $\hat{W}_{\vec\tau}$ depends on only $\mathbf E$ and $\vec \tau$, we can use the restriction \eqref{eqn:stress_ISRI_isotropy}. As we  have further assumed that the material supports a stress-free state, we need only use a simpler form (see the discussion leading to \eqref{eqn:stress_ISRI zero tau}), that is, we can take $\vec\tau_1=\mathbf{0}$ and write \eqref{eqn:stress_ISRI_isotropy} in terms of the quantities introduced in this section to obtain
\begin{equation} \label{eqn:strain_ISRI_tau}
\tcboxmath{
    \hat{W}_{\vec\tau}(\mathbf E_{12}, \mathbf 0) = J_1 \hat{W}_{\vec\tau}(\mathbf E, \sigmatauhat(\mathbf F_1, \mathbf 0)),
}
\end{equation}
for every $\mathbf F_1$ and $\mathbf F_2$, where $J_1$ is given by \eqref{eqn:J_1 expand}.  We can use the restriction \eqref{eqn:strain_ISRI_tau} to find the relationships between the constants. To do so, we first calculate the Cauchy stress in $\bb_2$ to be
\begin{equation}\label{eqn:tau to etwo}
\sigmatauhat(\mathbf F_1, \mathbf 0) =  
  [ 2\beta_1\tr \,\mathbf e  +  \beta_4\tr(\mathbf e^2) - (2\beta_1 - 3\beta_5) (\tr\,\mathbf e)^2] \mathbf I 
  + 2 [\beta_2 + (2\beta_1 - \beta_2 + \beta_4)\tr \,\mathbf e] \mathbf e   
  + (4\beta_2 + 3\beta_3) \mathbf e^2+O(\mathbf{e}^3).
\end{equation}
We then note that $\vec \tau = \sigmatauhat(\mathbf F_1, \mathbf 0)$ and invert the above to obtain
\begin{equation} \label{eqn:strain to stress}
	\mathbf e = [a_1 \tr \, \vec \tau + a_2 \tr(\vec \tau^2) +  a_3 \tr(\vec \tau)^2] \mathbf I + (a_4 + a_5 \tr \,\vec \tau) \vec \tau + a_6 \vec \tau^2 + \mathcal O(\vec \tau^3),
\end{equation}
where $a_j$, for $j=1,...,6$, are constants that can be written in terms of $\beta_i$, $i=1,..,5$ (see the supplementary Mathematica file for further  details). Substituting this expression into equations \eqref{eqn:tilde trE} gives the invariants of $\mathbf{E}_{12}$ in terms of the invariants of $\mathbf{E}$ and $\vec\tau$, which we  then use to write the left side of \eqref{eqn:strain_ISRI_tau} in terms of the mixed invariants of $\mathbf{E}$ and $\vec\tau$. For the right side of \eqref{eqn:strain_ISRI_tau} we note that $\vec \tau = \sigmatauhat(\mathbf F_1, \mathbf 0)$ so we can use the form \eqref{eqn:W tau}. Comparing the coefficients of each independent combination of invariants on both sides of \eqref{eqn:strain_ISRI_tau} leads  to
\begin{align} \notag
    & \beta_6 = 0,
    \quad \beta_7 = - \frac{\beta_1}{4\beta_2(3\beta_1+\beta_2)}, 
    \quad \beta_8 = \frac{1}{4\beta_2},
    \quad \beta_9 = -\frac{2\beta_2+\beta_3}{4\beta_2^{3}},
    \quad\beta_{10} =\frac{2\beta_1(4\beta_2+3\beta_3) + \beta_2(\beta_2-2\beta_4)}{8\beta_2^3(3\beta_1+\beta_2)}, 
\\ \notag
    & {\beta_{11} = -\frac{12\beta_1^3(\beta_2+\beta_3)+\beta_1^2\beta_2(7\beta_2+6\beta_3-6\beta_4)+\beta_1\beta_2^2(\beta_2-4\beta_4)+2\beta_2^3\beta_5}{8\beta_2^3(3\beta_1+\beta_2)^3},}\quad\beta_{12} = 2+\frac{3\beta_3}{2\beta_2},\notag
\end{align}
\begin{equation}\label{eqn:subISRI_1}   
\beta_{13} = - \frac{2\beta_1(2\beta_1+\beta_4) + \beta_2(\beta_1-3\beta_5)}{2\beta_2(3\beta_1 + \beta_2)}, 
\quad \beta _{14}=  \frac{2\beta_1 + \beta_4}{\beta_2},\quad\beta_{15}=  - \frac{\beta_1(4\beta_2 + 3\beta_3) + \beta_2(\beta_2-\beta_4)}{2\beta_2(3\beta_1+ \beta_2)}.
\end{equation}
This time, as required, we have 10 non-trivial equations, which, again, reduces the number of free parameters down to five. Again, if wished, we can set $\tilde{W}_\mathbf{E}(\mathbf{E}_{12})=\hat{W}_{\vec\tau}(\mathbf{E}_{12},\mathbf{0})$ to find
\begin{equation}\label{eqn:beta1to5}
    \beta_1=\frac{\lambda}{2},\quad\beta_2=\mu,\quad\beta_3=\frac{A}{3},\quad\beta_4=B,\quad\beta_5=\frac{C}{3}.
\end{equation}

For completeness, substituting \eqref{eqn:subISRI_1} into \eqref{eqn:W tau}, and relabelling the parameters as per equation~\eqref{eqn:beta1to5}, leads to
\begin{multline}\label{eqn:WEtau}
\hat{W}_{\vec\tau}(\mathbf E, \vec \tau) =
    \frac{ \lambda }{2} I_1^2 +  \frac{C}{3} I_1^3 + \mu I_2 + I_2 I_1 B + \frac{A}{3} I_3 -\frac{\lambda }{12 K \mu } I_4^2 + \frac{2 B K-\lambda (2 \lambda +\mu)}{36 K^2 \mu^2} I_4^3 
\\
     -\frac{ 3 A \lambda ^2 (\lambda +\mu )+4 \mu ^3 (B+C)}{162 K^3 \mu ^3} I_4^3 - \frac{2 \lambda  (B+\lambda )+\mu  (\lambda -2 C)}{6 K \mu } I_4 I_1^2  -\frac{A \lambda +2 \mu  (2 \lambda +\mu -B)}{6 K \mu } I_2 I_4  
\\
     +\frac{1}{4\mu } I_5  + \frac{A \lambda +\mu  (-2 B+4 \lambda +\mu )}{12 K \mu ^3} I_4 I_5  -\frac{A+6 \mu }{12 \mu ^3}I_6 +I_7 + \frac{B+\lambda }{\mu } I_7 I_1 + \frac{A}{2 \mu }I_9+2I_9,
\end{multline}
where $K = \lambda + 2 \mu/3$, which is the bulk modulus. 

The conclusion from this section is that expanding $\hat{W}_{\vec\tau}$ asymptotically in terms of its invariants, as shown in \eqref{eqn:W tau}, leads to too many elastic constants. Using the ISRI restriction \eqref{eqn:stress_ISRI_isotropy} makes $\Wtau$ equivalent to a classical third-order strain energy function.


\section{Conclusions}
\label{sec:concl}


This paper discusses how to build a sound nonlinear elastic constitutive theory when the reference configuration has a non-vanishing internal stress distribution. We considered stored energy functions of the form $\Wtau(\mathbf F, \vec \tau,\vec X)$ and $W(\mathbf F, \mathbf F_1,\vec X)$, where $\vec \tau$ is the initial stress, $\mathbf F_1$ is an initial elastic deformation gradient, $\mathbf F$ is the elastic deformation gradient from the stressed (or strained) reference configuration to the current configuration, and $\vec X$ is the position vector. We showed that $\Wtau(\mathbf F, \vec \tau,\vec X)$ and $W(\mathbf F, \mathbf F_1,\vec X)$ need to satisfy restrictions that are automatically satisfied by the classical strain energy functions. 

In \Cref{sec:elastic-spring}, we considered an initially stressed, linear elastic spring, and showed that, even in this simple case, restrictions are required on the stored energy function. This is necessary both for consistency with classical elasticity and for self-consistency. Stored energy functions that do not satisfy appropriate restrictions make contradictory predictions when used starting from different reference configurations.

In \Cref{sec:strained energies}, we deduced, in three spatial dimensions, restrictions for stored energies of the form $W(\mathbf F, \mathbf F_1,\vec X)$ from the assumptions of classical elasticity. In \Cref{sec:initially stress energies}, we derived restrictions on stored energy of the form $\Wtau(\mathbf F, \vec \tau,\vec X)$ from those derived in \Cref{sec:strained energies}. To do so, we further assumed that the stress could be inverted for the strain, which  holds for a wide range of constitutive models in classical elasticity. We gave an example of this stress inversion for third-order materials in \eqref{eqn:strain to stress}. A similar inversion is possible for an asymptotic expansion of any order. 

In \Cref{sec:third-order energies}, we give some examples to illustrate the theory when the strain and stress are small by considering third-order materials. We showed that, when unrestricted, stored energy functions of the form $\We(\mathbf E,\mathbf{e})$  and $\hat{W}_{\vec\tau}(\mathbf E,\vec\tau)$, asymptotically expanded up to third-order, have too many free parameters compared with classical third-order elastic materials \eqref{eqn:classical-third}, which have only five constants. We expect the number of constants to be the same, as $\We(\mathbf E,\mathbf{e})$, $\hat{W}_{\vec\tau}(\mathbf E,\vec\tau)$ and the classical third-order strain energy functions all account for the same quantities at the same asymptotic order. We showed that this inconsistency is resolved by enforcing the restrictions given in equations~\eqref{eqn:strain_ISRI_E} and~\eqref{eqn:strain_ISRI_tau}.


\section*{Supplementary Material}


The Mathematica file used to calculate the relationships between the constants in \Cref{sec:third-order energies} is provided as supplementary material.


\section*{Acknowledgement}


Artur Gower's contribution to this work was supported by the EPSRC grant EP/R014604/1. Pasquale Ciarletta's contribution to this work was supported by MUR grants Dipartimento di Eccellenza 2023-2027 and  PRIN 2020 2020F3NCPX. Tom Shearer (T.S.) and Michel Destrade (M.D.) would like to thank the Isaac Newton Institute for Mathematical Sciences for support and hospitality during the programme \textit{Uncertainty Quantification and Stochastic Modelling of Materials}, during which, much of their work on this paper was undertaken. They also thank the University of Galway for financial support which allowed T.S. to visit M.D. and continue working on the paper.

\appendix


\section{Alternative derivation of \textit{Initial Strain Reference Independence}} 
\label{sec:alt}


Similarly to the derivation in Section \ref{sec:restrict}, we could set the first argument of $W$ in equation~\eqref{def:W strain} to be the identity tensor $\mathbf I$ (i.e.\ $\mathbf{F}_2\rightarrow\mathbf{I}$) and the \textit{second} argument to be $\mathbf{F}_{12}$ (\i.e. $\mathbf{F}_1\rightarrow\mathbf{F}_{12}$), 
so that we are considering the case where \textit{all} of the deformation is initial deformation and there is no \textit{subsequent} deformation, which implies that $\vec X_2\rightarrow\vec x$, and, on the right side $J_1^{-1}\rightarrow J_{12}^{-1}$, $\mathbf{F}_2\mathbf{F}_1\rightarrow\mathbf{I}\mathbf{F}_{12}$ and $\phi_1^{-1}(\vec{X}_2)\rightarrow\phi_{12}^{-1}(\vec x)=\vec X_1$. Using these substitutions leads to
\begin{equation}
    W(\mathbf I, \mathbf F_{12}, \vec x)=W(\mathbf I, \mathbf F_2 \mathbf F_1, \vec x)=J_{12}^{-1}\tilde W(\mathbf I\mathbf F_{12},\vec X_1)=J_{12}^{-1}\tilde W(\mathbf F_2 \mathbf F_1,\vec X_1),
\end{equation}
and, comparing this with equations \eqref{def:W strain} and \eqref{eqn:strain_ISRI}, we obtain
\begin{equation} \label{eqn:strain_ISRI2}
	\tcboxmath{
W(\mathbf F_2 \mathbf F_1, \mathbf I, \vec X_1)=J_1 W(\mathbf F_2, \mathbf F_1, \vec X_2) = J_1J_2W(\mathbf I, \mathbf F_2 \mathbf F_1, \vec x),
}
\end{equation}
which must hold for every $\mathbf F_1$, $\mathbf F_2$, $\vec X_1$, $\vec X_2$ and $\vec x$. We remark that these three equalities are mathematically equivalent. 


\section{An inverse function for the stress}
\label{app:inverse stress}

To define a stored energy of the form $\Wtau(\mathbf F_2, \vec \tau_2,\vec X_2)$ in terms of a stored energy function of the form $W(\mathbf F_2, \mathbf F_1,\vec X_2)$, we need to invert the stress function $\tilde {\vec \sigma}$. Below, we demonstrate that this is possible for a wide class of functions $W$. Defining $\Wtau$ from a function $W$ has been previously carried out by Hoger \cite{hoger1997virtual}. In \Cref{sec:initial stress third-order}, we give an example of this stress inversion for third-order elasticity (see equation \eqref{eqn:strain to stress}). A similar inversion could be carried out for an asymptotic expansion of the stored energy function to any order.

The outline of our explanation is:
\begin{enumerate}
    \item We show the function $\tilde {\vec \sigma}$ is locally invertible for a large class of materials. That is, given $\mathbf U_1$, then, locally, around the point $\vec X_2$, where $\vec \tau_2 = \tilde {\vec \sigma}(\mathbf U_1, \vec X_1)$ there exists a unique inverse function $\tilde {\vec \sigma}^{-1}$ that satisfies $\tilde {\vec \sigma}^{-1}(\vec \tau_2, \vec X_2) = \mathbf U_1$.
    \item We let $\mathbf U_0$ be such that $\tilde W(\mathbf U_0, \vec X_1)$ is the global energy minimum. Then we show that there exists a unique global inverse $\tilde {\vec \sigma}^{-1}$ that passes through this energy minimum, that is, that satisfies $\tilde {\vec \sigma}^{-1}( \vec \tau_0, \vec X_1) = \mathbf U_0$, where $\tilde {\vec \sigma}(\mathbf U_0, \vec X_1) = \vec \tau_0$.
\end{enumerate}

\textbf{1. Local inversion.} For a wide class of materials, we expect $\tilde {\vec \sigma}$ to have a local inverse because local invertibility is a result of the material being stable under traction. We expect most materials to be stable for every $\mathbf U$ in some set $\mathcal U$ which we call the region of stability. 

To explain the connection between stability and a local inverse we note that: being stable under traction means that small changes in stress lead to small changes in strain, which implies that the partial derivative ${\partial \mathbf U}/{\partial {\vec \sigma}}$ is unique, exists, and is continuously differentiable \cite{antman1976fundamental,ball1976convexity}. The aforementioned properties of ${\partial \mathbf U}/{\partial {\vec \sigma}}$ also guarantee that there exists a local inverse of the function $\tilde{\vec \sigma}: \mathbf U \mapsto \vec \sigma$ for $\mathbf U \in \mathcal U$.

The above explains why we expect there to exist a local inverse for a large class of constitutive choices. However, to check if some given  $\tilde{\vec \sigma}$ is locally invertible, it is simpler to check whether: 
\begin{equation} \label{eqn:local stress invertibility}
	\Big\|\frac{\partial \tilde{\vec \sigma}(\mathbf U,\vec X_1)}{\partial \mathbf U} : \delta \mathbf U  \Big\| > c,
\end{equation}
for some positive real constant $c >0$, and any symmetric tensor $\delta \mathbf U$, where $\| \vec v \|$ is the Euclidean norm of any vector $\vec v$. Even for $c=0$, the condition \eqref{eqn:local stress invertibility} is enough to guarantee then there exists a local inverse, see \cite[Theorem 1.1.7.]{hormander2015analysis} and \cite{EthanInverse}. However, having $c>0$ allows us extend this inverse, as we do below.

\textbf{2. Defining a global inverse.} By using the condition~\eqref{eqn:local stress invertibility} for every $\mathbf U \in \mathcal U$ we can define an inverse $\tilde {\vec \sigma}^{-1}: \mathcal S \mapsto \mathcal U$ where the open set $\mathcal S := \{\tilde {\vec \sigma}(\mathbf{U},\vec X_1):\mathbf U \in \mathcal U\}$. To define this inverse uniquely, we need to choose one point, or state, for this inverse to pass through. That is, we need to choose a deformation $\mathbf U_0$ and a stress $\vec \tau_0$ such that $\tilde{\vec \sigma}(\mathbf U_0,\vec X_1) = \vec \tau_0$ and $\vec U_0 \in \mathcal U$. Often, a natural choice is $\mathbf U_0 = \mathbf I$ and $\vec \tau_0 = \tilde{\vec \sigma}(\mathbf I,\vec X_1)$, which is usually a stress-free state. However, to be more general, we choose:
\begin{equation}
\mathbf U_0 = \argmin_{\mathbf U} \tilde W(\mathbf U, \vec X_1).
\end{equation}

Now, we assume that there is an open set $\mathcal U$ around the point $\mathbf U_0$ such that the material is stable under traction for every $\mathbf U \in \mathcal U$, and, therefore,~\eqref{eqn:local stress invertibility} holds. As a consequence of \eqref{eqn:local stress invertibility}, there is a unique local inverse function $\tilde {\vec \sigma}^{-1}$ such that $\tilde {\vec \sigma}^{-1}(\vec \tau_0, \vec X_1) = \mathbf U_0$. We can uniquely extend this local inverse, by repeatedly using \eqref{eqn:local stress invertibility}, to define an inverse function $\tilde {\vec \sigma}^{-1} : \mathcal S \to \mathcal U$, 

In conclusion, $\tilde {\vec \sigma}^{-1}$ is an inverse of $\tilde {\vec \sigma}$ in the sense that
\begin{equation} \label{def:inverse sigma}
\tilde {\vec \sigma}^{-1}(\tilde {\vec \sigma}(\mathbf U,\vec X_1),\vec X_1) = \mathbf U \quad \text{and} \quad
\tilde {\vec \sigma}(\tilde {\vec \sigma}^{-1}(\vec \tau,\vec X_1),\vec X_1) = \vec \tau,
\end{equation}
for every $\mathbf U \in \mathcal U$ and $\vec \tau \in \mathcal S$, and $\tilde {\vec \sigma}^{-1}(\vec \tau_0, \vec X_1) = \mathbf U_0$.

\bibliographystyle{abbrv}
\bibliography{references} 

\end{document}